\documentclass{article}
\usepackage{amsmath}
\usepackage{amscd}
\usepackage{graphicx}
\usepackage{latexsym}
\usepackage{amsmath,amsthm,amsfonts,amssymb,cite}%
\usepackage{color}
\usepackage{latexsym}
\usepackage{amsmath}

\theoremstyle{remark}

\newtheorem*{rem}{Remark}
\renewcommand\dagger*
\renewcommand\Theta{\boldsymbol\theta}
\numberwithin{equation}{section}
\begin{document}
\hoffset = -2.4truecm \voffset = -2truecm
\renewcommand{\baselinestretch}{1.2}
\newcommand{\mb}{\makebox[10cm]{}\\ }
\title{Matrix integral solutions to the discrete KP hierarchy and its Pfaffianized version}
\author{St\'ephane Lafortune$^{1}$ and Chun-Xia Li$^{2,1}$\\
 \\$^1$ Department of Mathematics, College of Charleston, Charleston, SC 29401, USA
\\$^{2}$ School of Mathematical Sciences,
Capital Normal University,
Beijing 100048, CHINA }
\date{}
\maketitle

\begin{abstract}
Matrix integrals used in random matrix theory for the  study of eigenvalues of Hermitian ensembles have been shown to provide $\tau$-functions for several hierarchies of integrable equations.  In this article, we extend this relation by showing that such integrals can also provide $\tau$-functions for the discrete KP hierarchy and a coupled version of the same hierarchy obtained through the process of Pfaffianization. To do so, we consider the first equation of the discrete KP hierarchy, the Hirota-Miwa equation.  We write the Wronskian determinant solutions to the Hirota-Miwa equation and consider a particular form of matrix integrals, which we show is an example of those Wronskian solutions. The argument is then generalized to the whole hierarchy. A similar strategy is used for the Pfaffianized version of the hierarchy except that in that case, the solutions are written in terms of Pfaffians rather than determinants. 
\end{abstract}

\vskip 0.5cm

%Mathematics Subject Classification (2000). 37K10, 35Q51.
%
%\vskip 0.5cm
%
%PACS: 02.30Ik, 05.45Yv

\section{Introduction}

In random matrix theory, thanks to a trick introduced by Mehta \cite{Mehta81,Mehta04}, the matrix integrals used to compute the probability that the eigenvalues of the matrices in an Hermitian ensemble belong to a certain interval can be written as integrals over the eigenvalues. These integrals take the form
\begin{equation}
\idotsint \prod\limits_{1\leq j<k\leq N} |x_j-x_k|^{\beta} \exp\left(\sum_{j=1}^Nf(x_j)\right)  \,dx_1\, dx_2\,\dots\, dx_N,
\label{bintegral}
\end{equation}
where $\beta=1,2,$ or $4$ correspond, respectively, to the orthogonal, unitary, and symplectic ensembles and   $\exp\left(f(x)\right)$ is a weight function. Note that the terminology orthogonal, unitary, and symplectic for the different values of $\beta$ refers to the symmetry groups of the underlying measures  \cite{Tracy98,Mehta04}.

Integrals of the form  (\ref{bintegral})  have been studied in the context of integrable systems and matrix models in string theory \cite{Gerasimov91, Makeenko91,Adler95,AM1,ASM,DKM,DKM2,AM2,MMM,MM1,MM2,SK}. In this context, a dependence on the variables $t_i,\;i=1,2,3,\dots$ is introduced  in the following way
\begin{equation}
Z_N^{(\beta)}(t)={1\over N!}\int_{-\infty}^\infty\cdots\int_{-\infty}^\infty \prod\limits_{1\leq j<k\leq N} |x_j-x_k|^{\beta} 
 \exp\left(c\sum_{j=1}^{N}\eta(x_j,t)\right)  \,dx_1\, dx_2\,\dots\, dx_N,\;\;\eta(x,t)=\sum_{i=1}^{\infty}{x^it_i}+f(x),
\label{tintegral}
\end{equation} 
where the parameter $c$ is chosen such that $c=1$ for $\beta=1,2$ and $c=2$ for $\beta=4$. In the case $\beta=2$, the integral (\ref{tintegral}) gives the partition function for the Hermitian one-matrix model \cite{Gerasimov91,MMM,Makeenko91,AM2} and it was independently shown to provide a $\tau$-function for the  Toda \cite{Gerasimov91} and the KP \cite{Adler95,MM2} hierarchies. In the cases $\beta=1,4$ the integrals (\ref{tintegral}) provide $\tau$-functions for  the coupled KP hierarchy \cite{HO,SK} and the Pfaff Lattice   hierarchy \cite{ASM,Adler99}. While related through their bilinear forms, the KP and Pfaff hierarchies were discovered in different ways as the Pfaff lattice was first derived in \cite{ASM,Adler99} through Lie algebraic techniques, while the coupled KP equation was obtained  using the procedure now known as {\sl{Pfaffianization}} applied to the KP equation \cite{HO}. The procedure of Pfaffianization allows to generalize soliton equations with solutions represented by determinants to their coupled systems with solutions represented by Pfaffians.

 This connection of (\ref{bintegral}) with integrable systems was deepened by the authors of \cite{ZHL} who introduced dependence on the continuous variables $t_1,t_2,$ and $s$ and the discrete variables $n_1$ and $n_2$ in the following way
 \begin{equation}
\begin{aligned}\label{i1}
Z_N^{(\beta)}(t_1,t_2,n_1)=&{1\over N!}\int_{-1}^{\infty}\cdots\int_{-1}^{\infty}\prod\limits_{i=1}^N(1+x_i)^{c n_1} \prod\limits_{1\leq j<k\leq N} |x_j-x_k|^{\beta} \exp\left(\sum_{j=1}^Nc\eta(x_j,t)\right) dx_1\cdots dx_N,\\
Z_N^{(\beta)}(t_1,s,n_1)=&{1\over N!}\int_{0}^{\infty}\cdots\int_{0}^{\infty}\prod\limits_{i=1}^Nx_i^{cn_1} \prod\limits_{1\leq j<k\leq N} |x_j-x_k|^{\beta}  \exp\left(\sum_{j=1}^Nc\widehat{\eta}(x_j,t_1,s)\right) dx_1\cdots dx_N,\\
Z_N^{(\beta)}(t_1,n_1,n_2)=&{1\over N!}\int_{0}^{\infty}\cdots\int_{0}^{\infty}\prod\limits_{i=1}^Nx_i^{cn_1}(1+x_i)^{cn_2} \prod\limits_{1\leq j<k\leq N} |x_j-x_k|^{\beta}  \exp\left(\sum_{j=1}^Nc\overline{\eta}(x_j,t_1)\right) dx_1\cdots dx_N,
\end{aligned}
\end{equation}
where $\eta(x,t)=x t_1+x^2t_2+f(x),\;\;\widehat{\eta}(x,t_1,s)=xt_1-x^{-1}s+\widehat{f}(x),\;\;\overline{\eta}(x,t_1)=xt_1+\overline{f}(x)$, and where the parameter $c$ is chosen such that $c=1$ for $\beta=1,2$ and $c=2$ for $\beta=4$. The authors of \cite{ZHL}
 show that the matrix integrals in (\ref{i1}) with $\beta= 2$
 satisfy the bilinear versions of the differential-difference KP equation, the two-dimensional Toda lattice,
and the semi-discrete Toda lattice, respectively. Furthermore, in the cases $\beta=1$ and $\beta=4$,  the integrals in (\ref{i1}) are shown to satisfy the Pfaffianized versions of those equations. 

The most direct way that integrals of the form \eqref{tintegral} solve bilinear equations is to use the identities below, which were introduced in \cite{CA,NGB}:
\begin{align}
&{1\over N!}\idotsint\det\left[\phi_i(x_j)\right]_{1\le i, j\le N}\cdot\det\left[\psi_i(x_j)\right]_{1\le i, j\le N}\, \,dx_1\, dx_2\,\dots\, dx_N\nonumber\\
&=\det\left[\int\phi_i(x)\psi_j(x)\, dx\right]_{i,j=1,\dots,N},\label{ID1}\\
&{1\over N!}\idotsint\det[\phi_i(x_j)]_{i,j=1,\dots,N}\, \,dx_1\, dx_2\,\dots\, dx_N\nonumber\\
&=\mbox{Pf}\left[\,\,\iint\limits_{ x< y}(\phi_i(x)\,\phi_j(y)-\phi_i(y)\phi_j(x))\,dx\, dy\right]_{i,j=1,\cdots,N},\label{ID2}\\
&{1\over N!}\int\dots\int \mbox{det}[\phi_i(x_j),\psi_i(x_j)]_{i=1,\dots,2N,\,j=1,\dots,N}\,dx_1\dots dx_N\nonumber\\
&=\mbox{Pf}\left[\int (\phi_i(x)\psi_j(x)-\phi_j(x)\psi_i(x))dx\right]_{i,j=1,\dots,2N}.\label{ID3}
\end{align}
For the case $\beta=2$, the integrals in \eqref{tintegral} and \eqref{i1} are written as determinants using (\ref{ID1}), while in the case $\beta=1,4$, they are written 
as Pfaffians using \eqref{ID2} and \eqref{ID3}. As pointed out by Kakei in \cite{SK}, once written in that form, the integrals in \eqref{tintegral}  in the cases $\beta=1,2$ can be interpreted  as the continuum limits of $N$-soliton solutions. It can also be verified that the same observation applies to the integrals \eqref{i1} studied in \cite{ZHL}.  It is interesting to note that Tracy and Widom \cite{Tracy98} used the identities (\ref{ID1}), (\ref{ID2}), and \eqref{ID3}, to obtain Fredholm determinant expressions for expressions involved in the study of the spacings between consecutive eigenvalues.

%\begin{equation}

%In the cases $\beta=1,4$,  $Z_N^{(\beta)}(t)$ is a $\tau$-function for the coupled KP hierarchy \cite{SK,AM1,ASM,DKM2}. 
%\begin{aligned}
%\end{aligned}
%\end{equation} 

In this article, we will extend the relation between integrals of the form (\ref{bintegral}) and integrable systems by obtaining matrix integral solutions to the discrete KP hierarchy and its Pfaffianized version. This will be done by introducing a dependence of (\ref{bintegral}) on discrete variables and on an auxiliary continuous variable. We will be using the identities (\ref{ID1}), \eqref{ID2}, and (\ref{ID3}) in order to interpret our integrals as determinants and Pfaffians to prove our results. Once more, the case $\beta=2$ will give rise to solutions to the initial equations (in our case the discrete KP hierarchy), while the cases $\beta=1,4$ will give rise to solutions of the Pfaffianized versions.

Specifically, our strategy will be to consider the following integrals 
\begin{equation}
\label{gmi}
\begin{aligned}
Z_N^{(\beta)}(n_1,n_2,n_3,s)=&{1\over N!}\int_{-\infty}^{\frac{1}{a_3}}\cdots\int_{-\infty}^{\frac{1}{a_3}}\prod\limits_{i=1}^N(1-a_1x_i)^{-cn_1/a_1}(1-a_2x_i)^{-cn_2/a_2}\\
&(1-a_3x_i)^{-cn_3/a_3}\prod\limits_{1\le i<j\le N}|x_i-x_j|^{\beta} \exp{\left[c\sum_{i=1}^N\eta(x_i,s)\right]}dx_1\cdots dx_N,
\end{aligned}
\end{equation}
where $n_i,\,i=1,2,3$ are discrete variables, $s$ is a continuous variable,  the lattice parameters satisfy $0<a_1,a_2\le a_3$, $\eta(x,s)=xs+\eta_0(x)$, $c=1$ for $\beta=1,2$ and $c=2$ for $\beta=4$. The integrals \eqref{gmi} will provide 
solutions to the Hirota-Miwa and its Pfaffianized version. Those integrals will then be generalized (see \eqref{gmi2}) to obtained $\tau$-functions for the whole discrete KP hierarchy and its Pfaffianized version. 

The rest of this paper is organized as follows. %In Section 2, we will give a brief review on the matrix integral \eqref{tintegral} and the KP hierarchy and the coupled KP hierarchy. 
In Section \ref{2}, we present the Wronskian determinant solutions to the Hirota-Miwa equation and prove that the matrix integral \eqref{gmi} with $\beta=2$ is a particular example of such a solution. We then generalize 
the argument to the whole discrete KP hierarchy. 
In Section \ref{3}, we show that the integrals \eqref{gmi} with $\beta=1$ and $\beta=4$ give solutions to the Pfaffianized version of the Hirota-Miwa equation. 
Matrix integral solutions for the whole Pfaffianized version of the discrete KP hierarchy are also obtained.
Conclusions are given in Section \ref{c}.

\section{Matrix integral solutions to the discrete KP hierarchy}
\label{2}

\subsection{The  Hirota-Miwa equation and its determinant solutions}
\label{21}

The  Hirota-Miwa equation is the first bilinear equation of the discrete KP hierarchy. It is given by \cite{HR,MT}
\begin{align}\label{HM}
&a_1(a_2-a_3)\tau(n_1+a_1,n_2,n_3)\tau(n_1,n_2+a_2,n_3+a_3)\nonumber\\
&+a_2(a_3-a_1)\tau(n_1,n_2+a_2,n_3)\tau(n_1+a_1,n_2,n_3+a_3)\nonumber\\
&+a_3(a_1-a_2)\tau(n_1,n_2,n_3+a_3)\tau(n_1+a_1,n_2+a_2,n_3)=0,
\end{align}
where $a_1,a_3,a_3$ are lattice parameters and $\tau$ is a function of the discrete variables $n_1$, $n_2$ and $n_3$.  

In \cite{OHT}, the following Casorati determinant solution was obtained
\begin{align}\nonumber
\tau(n_1,n_2,n_3)=
\begin{vmatrix}\phi_1(n_1,n_2,n_3,0)&\phi_1(n_1,n_2,n_3,1)&\cdots&\phi_1(n_1,n_2,n_3,N-1)\\
\phi_2(n_1,n_2,n_3,0)&\phi_2(n_1,n_2,n_3,1)&\cdots&\phi_2(n_1,n_2,n_3,N-1)\\
\vdots&\vdots&\ddots&\vdots\\
\phi_N(n_1,n_2,n_3,0)&\phi_N(n_1,n_2,n_3,1)&\cdots&\phi_N(n_1,n_2,n_3,N-1)
\end{vmatrix},
\end{align}
where $\phi_i, i=1,\cdots,N$ satisfy the dispersion relations
\begin{equation}\label{dis}
\Delta_{n_j}\phi_i(n_1,n_2,n_3,s)={\phi_i(n_j)-\phi_i(n_j-a_j)\over a_j}=\phi_i(n_1,n_2,n_3,s+1),\,j=1,2,3.
\end{equation} 
Here $s$ is an auxiliary discrete variable.

For our purpose of obtaining matrix integral solutions, it will be more convenient to work with a Wronskian determinant solution. To find such a solution, we interpret $s$ as a continuous variable and 
write a determinant of a matrix whose columns are derivatives in the variable $s$.   More precisely, the expression for the Wronskian solution has the form 
\begin{align}\label{wds}
\tau(n_1,n_2,n_3)=
\begin{vmatrix}\phi_1^{(0)}(n_1,n_2,n_3,s)&\phi_1^{(1)}(n_1,n_2,n_3,s)&\cdots&\phi_1^{(N-1)}(n_1,n_2,n_3,s)\\
\phi_2^{(0)}(n_1,n_2,n_3,s)&\phi_2^{(1)}(n_1,n_2,n_3,s)&\cdots&\phi_2^{(N-1)}(n_1,n_2,n_3,s)\\
\vdots&\vdots&\ddots&\vdots\\
\phi_N^{(0)}(n_1,n_2,n_3,s)&\phi_N^{(1)}(n_1,n_2,n_3,s)&\cdots&\phi_N^{(N-1)}(n_1,n_2,n_3,s)
\end{vmatrix},
\end{align}
where $\phi_i^{(k)}(n_1,n_2,n_3,s)={\partial^k{\phi_i}\over \partial{s}^k}(n_1,n_2,n_3,s)$ and $\phi_i$ for $i=1,\cdots,N$ satisfy the dispersion relations
\begin{align}\label{DRW}
\Delta_{n_j}\phi_i(n_1,n_2,n_3,s)={\phi_i(n_j)-\phi_i(n_j-a_j)\over a_j}=\phi_i^{(1)}(n_1,n_2,n_3,s),\,\,j=1,2,3.
\end{align}

For completeness, we present the proof that the expression given in \eqref{wds} solves the Hirota-Miwa Equation. Here we adopt the compact notation introduced by Freeman and Nimmo \cite{FN1,FN2}. 
We denote the expression given in \eqref{wds} as
\begin{align}
\label{exptau}
\tau(n_1,n_2,n_3)=|0,1,\cdots,N-1|,
\end{align}
where each index $j$ for $j=0,\dots,N-1$ stands symbolically for a column vector given by 
\begin{equation*}
\begin{pmatrix}
\phi_1^{(j)}(n_1,n_2,n_3,s)\\
\phi_2^{(j)}(n_1,n_2,n_3,s)\\
\vdots\\
\phi_N^{(j)}(n_1,n_2,n_3,s)
\end{pmatrix}.
\end{equation*}
Based on the dispersion relations \eqref{DRW}, we have
\begin{align}
\nonumber
&\tau(n_i+a_i)=|0,1,\cdots,N-2,N-1_{n_i+a_i}|,\\
\nonumber&a_i\tau(n_i+a_i)=|0,1,\cdots,N-2,N-2_{n_i+a_i}|,\\
\nonumber&\tau(n_i+a_i,n_j+a_j)={1\over a_j-a_i}|0,1,\cdots,N-3,N-2_{n_i+a_i},N-2_{n_j+a_j}|,
\end{align}
where the notation $j_{n_i+a_i}$ is the column vector obtained by replacing $n_i$ with $n_i+a_i$ in the column vector corresponding to the index $j$. 
By substituting the expressions \eqref{exptau} into \eqref{HM}, the Hirota-Miwa equation is reduced to nothing but the Pl$\ddot{u}$cker relation \cite{HR2,HR1,HRN}
\begin{align}
&|0,1,\cdots,N-3,N-2,N-2_{n_1+a_1}||0,1,\cdots,N-3,N-2_{n_2+a_2},N-2_{n_3+a_3}|\nonumber\\
&-|0,1,\cdots,N-3,N-2,N-2_{n_2+a_2}||0,1,\cdots,N-3,N-2_{n_1+a_1},N-2_{n_3+a_3}|\nonumber\\
&+|0,1,\cdots,N-3,N-2,N-2_{n_3+a_3}||0,1,\cdots,N-3,N-2_{n_1+a_1},N-2_{n_2+a_2}|=0.\nonumber
\end{align}
%\begin{rem}
%Interpreting $s$ as a continuous variable is quite tricky but extremely important. It inspires us and makes it possible for us to design the proper matrix integral \eqref{gmi}. 
%\end{rem}

The $N$-soliton solution for equation \eqref{HM} obtained in  \cite{OHT} can be written as a Wronskian solution of the form \eqref{wds} with
\begin{align}\label{nss}
\phi_i(n_1,n_2,n_3,s)=\alpha_i\prod_{\nu=1}^3(1-p_ia_\nu)^{-n_\nu/a_\nu}\exp(p_is)+\beta_i\prod_{\nu=1}^3(1-q_ia_\nu)^{-n_\nu/a_\nu}\exp(q_is),
\end{align}
where $p_i$,\,$q_i$,\,$\alpha_i$,\,$\beta_i$ are arbitrary constants, which correspond to the wave numbers and phase parameters of the $i$-th soliton, respectively.

\subsection{Matrix integral solutions of the Hirota-Miwa equation}
\label{22}
Consider the case of $\beta=2$ for the integral $Z_N^{(\beta)}(n_1,n_2,n_3,s)$ given by \eqref{gmi} ,which we denote as $Z_N^{(\beta=2)}(n_1,n_2,n_3,s)$. In what follows, we show that  $Z_N^{(\beta=2)}(n_1,n_2,n_3,s)$ solves the Hirota-Miwa equation \eqref{HM} by showing that it is a particular case of the Wronskian solution \eqref{wds}.

Using the Vandermonde determinant
$$
\prod_{1\leq j<k\leq N}(x_j-x_k)=\det\left[x^{j-1}_k\right]_{j,k=1\cdots,N},
$$
we can express the integrand of $Z^{(\beta=2)}(n_1,n_2,n_3,s)$ as a product of two determinants
$$
\det\left[x_k^{j-1}\right]_{j,k=1\cdots,N}\cdot\det\left[x_k^{j-1}(1-a_1x_k)^{-n_1/a_1}(1-a_2x_k)^{-n_2/a_2}(1-a_3x_k)^{-n_3/a_3}\exp{\left(\eta(x_k,s)\right)}\right]_{j,k=1\cdots,N}.
$$ 
Then, using the identity \eqref{ID1}, the matrix integral $Z_N^{(\beta=2)}(n_1,n_2,n_3,s)$ can be written as
\begin{align}
Z_N^{(\beta=2)}(n_1,n_2,n_3,s)=&\det\left[\int_{-\infty}^{\frac{1}{a_3}} x^{i+j-2}(1-a_1x)^{-n_1/a_1}(1-a_2x)^{-n_2/a_2}(1-a_3x)^{-n_3/a_3}\exp{(xs+\eta_0(x))}dx\right]_{1\le i,j\le n}\nonumber\\
=&\begin{vmatrix}g_1^{(0)}(n_1,n_2,n_3,s)&g_1^{(1)}(n_1,n_2,n_3,s)&\cdots&g_1^{(N-1)}(n_1,n_2,n_3,s)\\
g_2^{(0)}(n_1,n_2,n_3,s)&g_2^{(1)}(n_1,n_2,n_3,s)&\cdots&g_2^{(N-1)}(n_1,n_2,n_3,s)\\
\vdots&\vdots&\ddots&\vdots\\
g_N^{(0)}(n_1,n_2,n_3,s)&g_N^{(1)}(n_1,n_2,n_3,s)&\cdots&g_N^{(N-1)}(n_1,n_2,n_3,s)\end{vmatrix},\nonumber
\end{align}
with 
\begin{equation}
\begin{aligned}
&g_i(n_1,n_2,n_3,s)=\int_{-\infty}^{1\over a_3}x^{i-1}(1-a_1x)^{-n_1/a_1}(1-a_2x)^{-n_2/a_2}(1-a_3x)^{-n_3/a_3}\exp{(xs+\eta_0(x))}dx,\\
&g_i^{(j)}(n_1,n_2,n_3,s)={\partial^j{g_i}\over\partial{s}^j}(n_1,n_2,n_3,s).
\end{aligned}
\label{gi}
\end{equation}

It is then a simple exercise to prove that the $g_i$ satisfy the dispersion relation \eqref{DRW}.
% i.e. 
%\begin{align}
%\Delta_{n_j}g_i(n_1,n_2,n_3,s)={g_i(n_j)-g_i(n_j-a_j)\over a_j}=g_i^{(1)}(n_1,n_2,n_3,s),\,\,j=1,2,3.
%\end{align}
The integral $Z_N^{(\beta=2)}(n_1,n_2,n_3,s)$ thus is a Wronskian solution to the Hirota-Miwa equation \eqref{HM}. 
Moreover, this solution can be seen as a continuum limit of the $N$-soliton solutions given by the Wronskian solution \eqref{wds} with $\phi_i$ as in \eqref{nss}.
Note that a similar interpretation applies to the whole discrete KP hierarchy treated below.

\subsection{Matrix integral solution of the discrete KP hierarchy}

The Hirota-Miwa equation is the first member of the discrete KP hierarchy \cite{OHT,DJM,KNW}. The Casorati determinant solutions of this hierarchy is presented in  \cite{OHT}. Just like we did in Section \ref{21}, we can rewrite it as a Wronskian solution, which takes the form
\begin{align}\label{wds2}
\tau({\bf{n}})=
\begin{vmatrix}\phi_1^{(0)}({\bf{n}},s)&\phi_1^{(1)}({\bf{n}},s)&\cdots&\phi_1^{(N-1)}({\bf{n}},s)\\
\phi_2^{(0)}({\bf{n}},s)&\phi_2^{(1)}({\bf{n}},s)&\cdots&\phi_2^{(N-1)}({\bf{n}},s)\\
\vdots&\vdots&\ddots&\vdots\\
\phi_N^{(0)}({\bf{n}},s)&\phi_N^{(1)}({\bf{n}},s)&\cdots&\phi_N^{(N-1)}({\bf{n}},s)
\end{vmatrix},
\end{align}
where ${\bf{n}}=(n_1,n_2,\dots,n_m)$, $\phi_i^{(k)}({\bf{n}},s)={\partial^k{\phi_i}\over \partial{s}^k}({\bf{n}},s)$ and $\phi_i$ for $i=1,\cdots,N$ satisfy the dispersion relations
\begin{align}\notag
\Delta_{n_j}\phi_i({\bf{n}},s)={\phi_i(n_j)-\phi_i(n_j-a_j)\over a_j}=\phi_i^{(1)}({\bf{n}},s),\,\,j=1,2,3,\dots,m,
\end{align} 
where $m\geq 3$. The case $m=3$ corresponds to the Hirota-Miwa equation treated in Section \ref{22}.
We now generalize the integral \eqref{gmi} so that it now depends on the discrete variables ${\bf{n}}=(n_1,n_2,\dots,n_m)$ in the following way
\begin{equation}
\label{gmi2}
\begin{aligned}
Z_N^{(\beta)}({\bf{n}},s)={1\over N!}\int_{-\infty}^{\frac{1}{a_m}}\cdots\int_{-\infty}^{\frac{1}{a_m}}\prod\limits_{i=1}^N\prod\limits_{k=1}^m(1-a_kx_i)^{-cn_k/a_k}\prod\limits_{1\le i<j\le N}|x_i-x_j|^{\beta} \exp{\left[\sum_{i=1}^Nc\eta(x_i,s)\right]}dx_1\cdots dx_N,
\end{aligned}
\end{equation}
where we assume that $a_m\geq a_j$ for $j=1,2,\dots m$. 
The argument used in Section \ref{22} is then easily generalized to show that the integral above in the case $\beta=2$ (with $c=1$) provides a solution of the form \eqref{wds2} for the discrete KP hierarchy.

\section{Matrix integral solutions to the Pfaffianized version of the discrete KP hierarchy}
\label{3}
In this section, we first recall some facts about Pfaffians and then present Pfaffian solutions and matrix integral solutions to the Pfaffianized Hirota-Miwa equation and the Pfaffianized version of the discrete KP hierarchy. 

As is known \cite{HR2,HR1,HRN,CER}, a Pfaffian $(1,2,\dots,2N)$ is defined recursively by  
\begin{align}\label{PD}
(1,2,\dots,2N)\triangleq\sum\limits_{j=2}^{2N}(-1)^j(1,j)(2,3,\dots,\hat{j},\dots,2N),
\end{align}
where $(i,j)=-(j,i)$ and $\hat{j}$ means that the index $j$ is omitted. For any given $2N\times 2N$ antisymmetric matrix $A_{2N}=[a_{ij}]_{1\le i,j\le 2N}$, the Pfaffian associated with $A_{2N}$ is defined as 
\begin{equation}\label{PM}\mbox{Pf}[A_{2N}]\triangleq(1,2,\dots,2N)\end{equation} with $(i,j)=-(j,i)=a_{ij}$. On the other hand, for any given Pfaffian $(1,2,\dots,2N)$, we can construct an antisymmetric matrix
\begin{equation*}A_{2N}=\begin{vmatrix}0&(1,2)&(1,3)&\cdots&(1,2N-1)&(1,2N)\\
-(1,2)&0&(2,3)&\cdots&(2,2N-1)&(2,2N)\\
\vdots&\vdots&\vdots&\ddots&\vdots&\vdots\\
-(1,2N-1)&-(2,2N-1)&-(3,2N-1)&\cdots&0&-(2N-1,2N)\\
-(1,2N)&-(2,2N)&-(3,2N)&\cdots&-(2N-1,2N)&0
\end{vmatrix}\end{equation*}
whose Pfaffian is $\mbox{Pf}[A_{2N}]=(1,2,\dots,2N)$ by the definition \eqref{PD}.

%A Pfaffian $(1,2,\dots,2N)$ is defined recursively by  \cite{HR2,HR1,HRN,CER}
%\begin{align*}
%(1,2,\dots,2N)=\sum\limits_{j=2}^{2N}(-1)^j(1,j)(2,3,\dots,\hat{j},\dots,2N),
%\end{align*}
%where $(i,j)=-(j,i)$ and $\hat{j}$ means that the index $j$ is omitted. Under this definition, given any $2N\times 2N$ antisymmetric matrix $A_{2N}=[a_{ij}]_{1\le i,j\le 2N}$, the Pfaffian associated with $A_{2N}$ can be denoted as 
%$$
%\mbox{Pf}\left[A_{2N}\right]=(1,2,\dots,2N)
%$$
%with $(i,j)=-(j,i)=a_{ij}$.}
\subsection{Pfaffianized version of the Hirota-Miwa}
\label{31}

The Pfaffianized version of the Hirota-Miwa equation was obtained in \cite{GNT} using the procedure of Pfaffianization proposed by Hirota and Ohta \cite{HO}. It takes the form of the following system of three coupled equations
\begin{equation}
\begin{aligned}
\label{CHM}
&a_1a_2a_3\left[a_1(a_2-a_3)\tau(n_1-a_1,n_2,n_3)\tau(n_1,n_2-a_2,n_3-a_3)\right.\\
&+a_2(a_3-a_1)\tau(n_1,n_2-a_2,n_3)\tau(n_1-a_1,n_2,n_3-a_3)\\
&\left.+a_3(a_1-a_2)\tau(n_1,n_2,n_3-a_3)\tau(n_1-a_1,n_2-a_2,n_3)\right]\\
&+(a_1-a_2)(a_2-a_3)(a_3-a_1)\sigma(n_1,n_2,n_3)\rho(n_1-a_1,n_2-a_2,n_3-a_3)=0,\\ 
&a_1a_2a_3\left[a_1(a_2-a_3)\sigma(n_1-a_1,n_2,n_3)\tau(n_1,n_2-a_2,n_3-a_3)\right.\\
&+a_2(a_3-a_1)\sigma(n_1,n_2-a_2,n_3)\tau(n_1-a_1,n_2,n_3-a_3)\\
&\left.+a_3(a_1-a_2)\sigma(n_1,n_2,n_3-a_3)\tau(n_1-a_1,n_2-a_2,n_3)\right]\\
&+(a_1-a_2)(a_2-a_3)(a_3-a_1)\sigma(n_1,n_2,n_3)\tau(n_1-a_1,n_2-a_2,n_3-a_3)=0,\\
&a_1a_2a_3\left[a_1(a_2-a_3)\tau(n_1-a_1,n_2,n_3)\rho(n_1,n_2-a_2,n_3-a_3)\right.\\
&+a_2(a_3-a_1)\tau(n_1,n_2-a_2,n_3)\rho(n_1-a_1,n_2,n_3-a_3)\\
&\left.+a_3(a_1-a_2)\tau(n_1,n_2,n_3-a_3)\rho(n_1-a_1,n_2-a_2,n_3)\right]\\
&+(a_1-a_2)(a_2-a_3)(a_3-a_1)\tau(n_1,n_2,n_3)\rho(n_1-a_1,n_2-a_2,n_3-a_3)=0.
\end{aligned}
\end{equation}

The system above has the Pfaffian solutions
\begin{align}
&\tau=\tau(n_1,n_2,n_3)=(1,2,\cdots,N), \, N\,\, \mbox{even},\nonumber\\
&\sigma=\sigma(n_1,n_2,n_3)=(1,2,\cdots,N,N+1,N+2),\label{tsol}\\
&\rho=\rho(n_1,n_2,n_3)=(1,2,\cdots,N-2),\nonumber
\end{align}
where the Pfaffian entries  $(i,j)$ in the above-mentioned three Pfaffians are functions of the discrete variables $n_1,\,n_2$ and $n_3$ which satisfy the following dispersion relations
\begin{align}\label{PDR}
(i,j)_{n_k-a_k}=(i,j)-a_k(i+1,j)-a_k(i,j+1)+a_k^2(i+1,j+1),\,\,k=1,2,3.
\end{align}
In \eqref{PDR}, the notation $(i,j)_{n_k-a_k}$ represents the operation on $(i,j)$ which replaces the discrete variable $n_k$ in $(i,j)$ by $n_k-a_k$. The terms on the right hand side of \eqref{PDR} are various Pfaffian entries. Based on the definition of a Pfaffian \eqref{PD}, once we know the dispersion relations \eqref{PDR} that the Pfaffian entries $(i,j)$ satisfy for $i,j=1,2,\dots,N+2$,  it is enough to prove that $\tau,\sigma$ and $\rho$ given by \eqref{tsol} are Pfaffian solutions to the Pfaffianized version of the Hirota-Miwa equation \eqref{CHM}. As for the antisymmetric matrices associated to the three Pfaffians in \eqref{tsol}, they are related but not the same. For example, if we assume that $A_{N-2},\, A_{N}$ and $A_{N+2}$ are three antisymmetric matrices such that $\rho=\mbox{Pf}[A_{N-2}], \,\tau=\mbox{Pf}[A_{N}]$ and $\sigma=\mbox{Pf}[A_{N+2}]$, respectively, then it is obvious that $A_{N+2}$ is an enlarged matrix of $A_{N}$ and $A_N$ is an enlarged matrix of $A_{N-2}$.

%The Pfaffian solutions can be proved 
%\begin{align}
%&\tau(n_i-a_i)=\mbox{Pf}(1,\cdots,N+1,c_i),\\
%&\tau(n_i-a_i,n_j-a_j)={a_ia_j\over a_j-a_i}\mbox{Pf}(1,\cdots,N+1,N+2,c_i,c_j),\\
%&\tau(n_1-a_1,n_2-a_2,n_3-a_3)={a_1^2a_2^2a_3^2\over (a_1-a_2)(a_2-a_3)(a_3-a_1)}\mbox{Pf}(1,\cdots,N+1,N+2,N+3,c_1,c_2,c_3).
%\end{align}
%where $\mbox{Pf}(i,c_j)=a_j^{N+1-i}$.

We would like to point out that an example of the Pfaffian entry $(i,j)$ which satisfies \eqref{PDR} is
\begin{align}\label{sol}
(i,j)=\sum\limits_{k=1}^N \left(p_k^{i-1}q_k^{j-1}-q_k^{i-1}p_k^{j-1}\right)\prod_{\nu=1}^3{[(1-a_\nu p_k)(1-a_\nu q_k)]}^{-n_\nu/a_v}\exp[\eta(p_k,s)+\eta(q_k,s)],
\end{align}
where $p_k$, and $q_k$ are arbitrary constants. 
Under this choice, the Pfaffian solution $\tau=(1,2,\dots,N)$ gives the $(N/2)$-soliton solutions to the Pfaffianized version of the Hirota-Miwa equation.

\begin{rem}
By refering to  \cite{HO,HR2,HR1}, we know that the coupled KP equation has Wronski-type pfaffian solutions $\tau=(0,1,\cdots,2N-1),\, \sigma=(0,1,\dots,2N-3)$ and $\hat{\sigma}=(0,1,\dots,2N+1)$ with $(l,m)$ satisfying the dispersion relation ${\partial\over\partial x_n}(l,m)=(l+n,m)+(l,m+n)$ where $x_1=x,x_2=y,x_3=t$. One such an example is   $(l,m)=\sum\limits_{k=1}^{2N}\left[\Phi_k^{(l-1)}\Psi_k^{(m-1)}-\Phi_k^{(m-1)}\Psi_k^{(l-1)}\right]$, where $\Phi_k^{(l)}$ and $\Psi_k^{(l)}$ stand for the $l$-th derivatives with respect to $x_1$ of functions $\Phi_k$ and $\Psi_k$ satisfying ${\partial\over\partial x_n}\Phi_k=\Phi_k^{(n)}$ and ${\partial\over\partial x_n}\Psi_k=\Psi_k^{(n)}$. Further on, if we choose $\Phi_k=\exp[p_kx+p_k^2y+p_k^3t]$ and $\Psi_k=\exp[q_kx+q_k^2+q_k^3t]$, then $(l,m)=\sum\limits_{k=1}^{2N}\left[p_k^{l-1}q_k^{m-1}-q_k^{l-1}p_k^{m-1}\right]\exp[(p_k+q_k)x+(p_k^2+q_k^2)y+(p_k^3+q_k^3)t]$ coincides with the expression $a_{l,m}$ given in \cite{SK} and can be used to generate multi-solition solutions to the coupled KP equation given in \cite{IWS}. Along this line, we can derive multi-soliton solutions to the coupled Hirota-Miwa equation with the pfaffian entry $(i,j)$ given by \eqref{sol} in the same way.
\end{rem}

\paragraph{Matrix integral solutions (I).}
Consider the case $\beta=1$. By virtue of the identity \eqref{ID2}, the matrix integral \eqref{gmi} turns into
\begin{align}\nonumber
&Z_N^{(\beta=1)}(n_1,n_2,n_3,s)\nonumber\\
&={1\over N!}\int_{-\infty}^{\frac{1}{a_3}}\cdots\int_{-\infty}^{\frac{1}{a_3}}\det\left[x_j^{i-1}(1-a_1x_j)^{-n_1/a_1}(1-a_2x_j)^{-n_2/a_2}(1-a_3x_j)^{-n_3/a_3}\exp{\left(\eta(x_j,s)\right)}\right]_{i,j=1\cdots,N} \nonumber\\
&=\mbox{Pf}\left[\iint\limits_{x<y} (x^{i-1}y^{j-1}-y^{i-1}x^{j-1})[(1-a_1x)(1-a_1y)]^{-n_1/a_1}\right.\nonumber\\
&\quad\left.[(1-a_2x)(1-a_2y)]^{-n_2/a_2}[(1-a_3x)(1-a_3y)]^{-n_3/a_3}\exp^{\eta(x,s)+\eta(y,s)}dx\,dy\right]_{i,j=1,\cdots,N},\nonumber
\end{align}
whose Pfaffian entry $(i,j)$ is given by
\begin{equation}
\begin{aligned}\label{pe1}
(i,j)=&\iint\limits_{x<y} (x^{i-1}y^{j-1}-y^{i-1}x^{j-1})[(1-a_1x)(1-a_1y)]^{-n_1/a_1}\\
&[(1-a_2x)(1-a_2y)]^{-n_2/a_2}(1-a_3x)(1-a_3y)]^{-n_3/a_3}\exp^{\eta(x,s)+\eta(y,s)}dx\,dy.
\end{aligned}
\end{equation}
It is easy to check that the Pfaffian entry \eqref{pe1} satisfies the dispersion relation \eqref{PDR}. Therefore, the matrix integrals $\tau=Z_N^{(\beta=1)}(n_1,n_2,n_3,s)$, $\sigma=Z_{N+2}^{(\beta=1)}(n_1,n_2,n_3,s)$ together with $\rho=Z_{N-2}^{(\beta=1)}(n_1,n_2,n_3,s)$ provide solutions to  the coupled Hirota-Miwa system \eqref{CHM}. Moreover, the Pfaffian solutions obtained with \eqref{pe1} can be interpreted as a continuum of the multisoliton Pfaffian solutions defined by  \eqref{sol}.

\paragraph{Matrix integral solutions (II).} With regard to the twofold Vandermonde determinant, the following identity was presented in \cite{Tracy98}
\begin{align}\label{TV}
\prod\limits_{1\leq j<k\leq N}(x_j-x_k)^4=\det\left[x_k^j,(j-1)x_k^j\right]_{j=1,\cdots,2N,\,k=1,\cdots,N}.
\end{align}
Consider the case $\beta=4$. By resorting to the identities \eqref{ID3} and \eqref{TV}, the matrix integral \eqref{gmi} becomes
\begin{align}\nonumber
&Z_N^{(\beta=4)}(n_1,n_2,n_3,s)\nonumber\\
&={1\over N!}\int_{-\infty}^{\frac{1}{a_3}}\cdots\int_{-\infty}^{\frac{1}{a_3}}\det\left[x_j^i(1-a_1x_j)^{-n_1/a_1}(1-a_2x_j)^{-n_2/a_2}(1-a_3x_j)^{-n_3/a_3}\exp(\eta(x_j,s)),\right.\nonumber\\
&\quad\left.(i-1)x_j^i(1-a_1x_j)^{-n_1/a_1}(1-a_2x_j)^{-n_2/a_2}(1-a_3x_j)^{-n_3/a_3}\exp(\eta(x_j,s))\right]dx_1\cdots dx_N\nonumber\\
\nonumber&=\mbox{Pf}\left[\int_{-\infty}^{\frac{1}{a_3}}(j-i)x^{i+j-3}(1-a_1x)^{-2n_1/a_1}(1-a_2x)^{-2n_2/a_2}(1-a_3x)^{-2n_3/a_3}\exp(2\eta(x,s))dx\right]_{i,j=1,\cdots,2N},
\end{align}
whose Pfaffian entry $(i,j)$ entry is given by
\begin{align}
(i,j)=\int_{-\infty}^{\frac{1}{a_3}}(j-i)x^{i+j-3}(1-a_1x)^{-2n_1/a_1}(1-a_2x)^{-2n_2/a_2}(1-a_3x)^{-2n_3/a_3}\exp(2\eta(x,s))dx.\label{pe2}
\end{align}
Similarly, one can prove that the Pfaffian entry \eqref{pe2} satisfies the dispersion relation \eqref{PDR} as well. Thus, the matrix integrals $\tau=Z_N^{(\beta=1)}(n_1,n_2,n_3,s),\, \sigma=Z_{N+1}^{(\beta=1)}(n_1,n_2,n_3,s)$ as well as $\rho=Z_{N-1}^{(\beta=1)}(n_1,n_2,n_3,s)$ give solutions to the coupled Hirota-Miwa system \eqref{CHM}.

\subsection{Pfaffianized version of the discrete KP hierarchy}

The Pfaffianized version of the whole discrete KP hierarchy and its Pfaffian solutions were derived in \cite{Ohta04}. The Pfaffian solutions take the form
\begin{equation}
\begin{aligned}\label{ns}
&\tau=\tau({\bf{n}})=(1,2,\cdots,N), \, N\,\, \mbox{even},\\
&\sigma=\sigma({\bf{n}})=(1,2,\cdots,N,N+1,N+2),\\
&\rho=\rho({\bf{n}})=(1,2,\cdots,N-2),
\end{aligned}
\end{equation}
where ${\bf{n}}=(n_1,n_2,\dots,n_m)$ and
where the Pfaffian entry $(i,j)$ satisfies
\begin{align}\notag
(i,j)_{n_k-a_k}=(i,j)-a_k(i+1,j)-a_k(i,j+1)+a_k^2(i+1,j+1),\,\,k=1,2,\dots m.
\end{align}
If we now consider the general integral \eqref{gmi2}, the argument used in Section \ref{31} is then easily generalized to show that the integral above in the cases $\beta=1$ (with $c=1$) and $\beta=4$ (with $c=2$)  provide a solution of the form \eqref{ns} for the Pfaffianized discrete KP hierarchy. 

\section{Conclusion}
\label{c}
We have considered matrix integrals used in certain probability computations for  Hermitian ensembles. Such matrices have been shown before to be useful in the context of integrable systems, string theory, and random matrix theory by providing partition functions and solutions to integrable equations. More precisely, by inserting an appropriate dependence on various variables in the integrals \eqref{bintegral}, one can obtain $\tau$-functions for a number of integrable equations, together with their Pfaffianized versions \cite{ASM,Adler99,ZHL,HO,SK}.  In all the examples studied in the literature, the case $\beta=2$ relates to a given integrable equation, while the cases $\beta=1,4$ relates to the Pfaffianized version of the same  equation.  Furthermore, the solutions obtained in this way can often be interpreted as a continuous limit of multi-soliton solutions. 

In this article, we have shown that integrals of the form \eqref{bintegral} can provide $\tau$-functions for the discrete KP hierarchy and its Pfaffianized version by introducing an appropriate dependence on the discrete variables. More precisely, we have considered the integrals \eqref{gmi} and \eqref{gmi2} and showed that in the case $\beta=2$, they provide solutions to the discrete KP hierarchy (starting with the Hirota-Miwa equation \eqref{HM}), while the cases $\beta=1,4$ provide solutions for the Pfaffianized version of the discrete KP hierarchy (starting with the Pfaffianized Hirota-Miwa Equation given in \eqref{CHM}).

\section*{Acknowledgement}
This work was supported by the National Natural Science Foundation of China (Grant No. 11271266 and 11371323), Beijing Natural Science Foundation (Grant No. 1162003) and the China Scholarship Council. The authors would like to thank Professor Xing-Biao Hu, Alex Kasman, and Virgil Pierce for valuable discussions. Dr. Li would also like to thank for the hospitality of the Department of Mathematics during her visit to the College of Charleston.

\end{document}